\documentclass{PoS}

\title{Mixing of S-Wave Charmonia with $\mathrm{D}\overline{\mathrm{D}}$ Molecule States}

\ShortTitle{Mixing of S-wave charmonia with
$\mathrm{D}\overline{\mathrm{D}}$ molecule states}

\author{\speaker{C.~Ehmann}, Gunnar Bali\\
        Institut f\"ur Theoretische Physik, Universit\"at Regensburg\\
93040 Regensburg, Germany.\\
        E-mail:\\
\email{christian.ehmann@physik.uni-regensburg.de}
\email{gunnar.bali@physik.uni-regensburg.de}}

\abstract{Charmonium states can decay into pairs of
$D$ and $\overline{D}$ mesons if their masses are above the
allowed decay thresholds. In general $c\bar{c}$ states near threshold
will also undergo mixing with
$D\overline{D}$ molecular (or tetraquark) states,
by creation and annihilation of light quark-antiquark pairs.
The investigation of such effects sheds light on the higher
Fock state contributions to charmonium wavefunctions
and on mass shifts, relative to a scenario where such mixing effects
are neglected.
A variational approach is applied to a mixing matrix
between operators of both sectors, of $c\bar{c}$ and of $D\overline{D}$
molecular type.
The efficient calculation of several diagrams appearing in this matrix
requires all-to-all propagators, which are realized by
sophisticated stochastic estimator techniques.
The runs are performed on $n_F=2$ $24^3\times 48$ lattice
volumes with $m_{\pi} \approx 280$ MeV, using the
non-perturbatively improved clover Wilson action, both
for valence and for sea quarks. 
}

\FullConference{The XXVII International Symposium on Lattice Field Theory - LAT2009\\
		 July 26-31 2009\\
		 Peking University, Beijing, China}

\begin{document}

\section{Introduction}
Nonperturbative simulations have demonstrated that quantum chromodynamics
(QCD) predicts the confinement of objects that carry a color charge.
The hadronic color singlet states with baryon number $B=0$ are called
mesons. Within the nonrelativistic quark model these 
contain exactly one quark and one antiquark. However, at least in principle,
QCD offers the possibility to construct such states entirely out of gluons
(glueballs), out of quark, antiquark and gluons (hybrids), out
of two quarks and two antiquarks (molecules or tetraquarks) and even
higher Fock states. In general these states will undergo mixing
and not all these states will be stable against strong decays.

The charmonium sector provides us with a particularly rich
laboratory for the study of these effects. Several experimentally
discovered charmonium states are suspected to contain large contributions
from non-$c\bar{c}$ configurations \cite{Barnes:2003vb,Godfrey:2008nc,Bignamini:2009sk}. Here we wish to address the contribution of
$c\bar{q}q\bar{c}$ molecules\footnote{We use the term ``molecule'' synonymous
also for tetraquarks.} to S-wave charmonia
and of $c\bar{c}$ states to $\mathrm{D}\overline{\mathrm{D}}$
bound states.

QCD predicts the spectrum within a sector of fixed $J^{PC}$, isospin $I$,
strangeness $S$ and charm $C$. Classifying the resulting
states according to their (presumed) quark content will be model
dependent. However, there are examples where clearly one Fock state
dominates over another and where mixing is small. For instance the
($J^{PC}=0^{-+}$, $I=S=C=0$) $\eta_c$ state is clearly different
from the $\eta'$ state with exactly the same quantum numbers
and in this case dynamic
mixing effects turn out to be negligible~\cite{Ehmann:2009ki},
such that we can identify the leading Fock contribution to
the former state as $c\bar{c}$ while the $\eta'$ is composed
of strange and light quark-antiquark pairs.

With such caveats in mind, we will somewhat carelessly
use the term ``$\eta_c$-$D_1\overline{D}^*$-mixing'' when
we mean mixing between states that couple to $c\gamma_5\bar{c}$-type
operators and states created by, $(c\gamma_5\gamma_i\bar{q})(\bar{c}\gamma_iq)$.
These ``unperturbed'' states can be accessed on the lattice by appropriate
interpolating fields. Their mixing is studied by applying a
variational approach. One of the problems is that, unless the
two sectors completely decouple from each other, in the limit of large
Euclidean times the states created by all these operators will
decay into the same ground state: the ground state created by
the $c\bar{c}$ type operators, that we call the unmixed
$\eta_c$, will intrinsically already contain a $D_1\overline{D}^*$
contribution.
However, if the perturbative approach that we outline
below is justified, then this implicit mixing~\cite{Bali:2005fu}
will first appear at second order  in an
expansion parameter $\lambda$, while the transition matrix element
between the two sectors (explicit mixing)
will be of order $\lambda$. Consequently,
at intermediate Euclidean times implicit mixing might be negligible,
reducing the model dependence of our ansatz.

Our perturbative set-up is as follows.
We expect the physical $\eta_c$ wavefunction
at first order in a parameter $\lambda$ to read,
\begin{equation}
  |\eta_c\rangle = \frac{1}{\mathcal{N}}\left(|c\bar{c}\rangle + \lambda \frac{\langle c\bar{q}q\bar{c}| H_1 | c\bar{c} \rangle}{E(c\bar{c})-E(c\bar{q}q\bar{c})}| c\bar{q}q\bar{c}\rangle\right)\,,
\end{equation}
with a normalisation factor $\mathcal{N}$ and a
(hopefully) small\footnote{In some cases $\lambda$ might be large
but we should be able to detect this through implicit mixing, i.e.\ decay
of the state created by the $c\bar{q}q\bar{c}$ operators into the
$c\bar{c}$ states at small Euclidean times.}
coupling constant $\lambda$ appearing in the mixing
vertex of the Hamiltonian $H_1$. While we do not know the functional
form of $H_1$ or of the unperturbed wavefunctions, we can evaluate
all the relevant matrix elements on the lattice.
One important thing to note here is the dependence of the mixing
on the light quark mass. With decreasing $m_q$, the denominator
obviously becomes smaller, but the mixing matrix element in the
numerator is expected to increase, since the probability
for creating a light quark-antiquark pair should be inversely dependent
on the light quark mass. Therefore we expect mixing effects to
increase at smaller light quark masses. 

\FIGURE
{\includegraphics[width=.9\textwidth,clip]{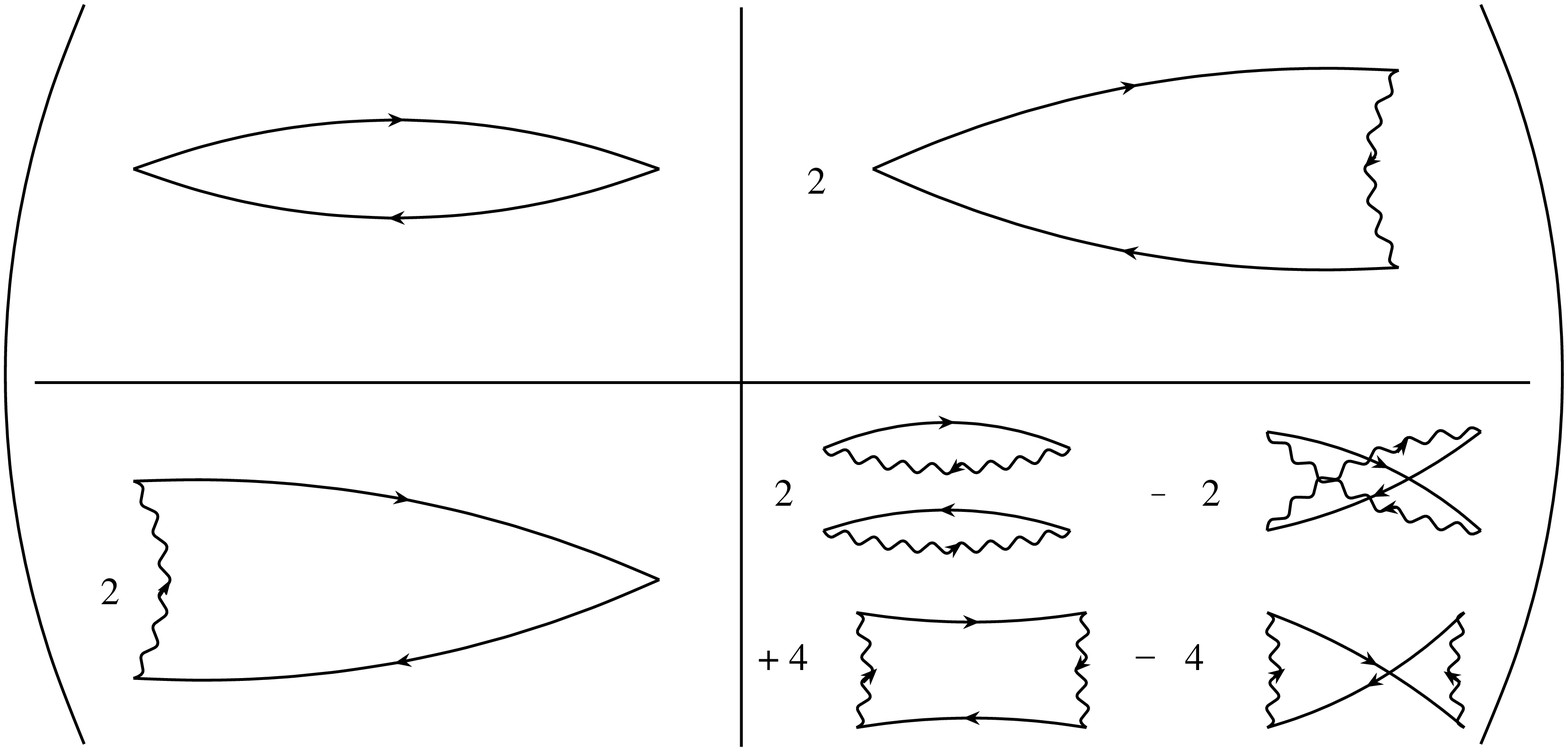}
\caption{The mixing matrix. Solid lines represent charm quarks, wiggled
lines light quarks.\label{mixingmatrix_fig}}}
\section{Simulation}
We aim to calculate the coefficients in the expansion of the
QCD eigenstates into the trial interpolating fields nonperturbatively,
by diagonalizing a matrix of cross correlators including both $c\bar{c}$
and molecular operators. For each type of operator we apply three
different types of smearing: local, narrow and wide, where these terms
indicate the number of Wuppertal smearing steps with $\delta=0.3$,
employing spatial APE smeared ($n_{\rm APE}=15$, $\alpha=0.3$) parallel
transporters to smooth the trial wavefunctions. The number of fermion
field smearing iterations is determined by optimizing the effective
masses separately for each of the two sectors.

In figure~\ref{mixingmatrix_fig} we sketch the structure of the
mixing matrix. The 
different smearing levels are omitted for the sake of clarity.
Solid lines represent charm quark propagators and wiggly lines light quark
propagators. The prefactors are due to the two mass degenerate light
sea quark flavors. The upper left corner contains the $c\bar{c}$,
the lower right corner the molecular sector. Nonvanishing off-diagonal elements
indicate mixing. The spatial separation within the molecular
operators was tuned by maximizing the magnitude of the off-diagonal
element. The optimal value was $r=4a\approx 0.3\,$fm.

The charm-anticharm annihilation diagrams were
omitted in the present study where we focus on $c\bar{c}$-molecule
mixing near threshold. We studied these previously
in the context of $\eta_c$-$\eta'$ mixing where their effect turned out
to be negligible~\cite{Ehmann:2009ki}.

For the evaluation of the last two diagrams of the molecular sector, light
all-to-all propagators are necessary. $O(\sim100)$ complex $\mathbb{Z}_2$
stochastic estimates per configuration were calculated
for this purpose, with the application of sophisticated noise reduction
methods like staggered-spin-color dilution and hopping parameter
acceleration as already used in previous studies
\cite{Bali:2008sx,Ehmann:2009ki,Bali:2009hu}.

\TABULAR{|c|c|c|c|c|}
{ \hline
 $J^{PC}$ & $\Gamma_M$ & $\Gamma_Y^1$ & $\Gamma_Y^2$ & $s$ \\
 \hline
 $0^{-+}$ & $\gamma_5$ & $\gamma_i$ & $\gamma_i\gamma_5$ & $0$ \\
 $1^{--}$ & $\gamma_i$ & $\gamma_5$ & $\gamma_i\gamma_5$ & $1$ \\
 $1^{++}$ & $\gamma_i\gamma_5$ & $\gamma_5$ & $\gamma_i$ & $1$ \\
\hline}
{$\Gamma$ structures of meson and molecular interpolating fields.\label{op_tab}}

The generic form of our meson interpolators, centred around a position
$x$, reads,
\begin{equation}
M(x)=(\bar{c}\Gamma_Mc)_x \, ,
\end{equation}
while the molecular interpolators with separation $r$ look like,
\begin{equation}
Y(x,r)=\frac{1}{\sqrt{2}}\left((\bar{q}\Gamma_Y^1c)_x(\bar{c}\Gamma_Y^2q)_{x+r} + (-)^s (\bar{c}\Gamma_Y^1q)_x(\bar{q}\Gamma_Y^2c)_{x+r}\right) \, .
\end{equation}
In table~\ref{op_tab} we display the explicit $\Gamma$ structures for
the $J^{PC}=0^{-+},1^{--}$ and $1^{++}$ channels.

The variational method consists of
solving a generalized eigenvalue problem,
\begin{equation}
C(t_0)^{-1/2} C(t) C(t_0)^{-1/2} \psi^{\alpha}
=
\lambda^{\alpha}(t,t_0) \psi^{\alpha}\,,
\end{equation}
see refs.~\cite{Ehmann:2007hj,Blossier:2009kd,Mi85}
for details. For sufficiently large times the eigenvalues
and -vectors will approach their asymptotic values. The components
of a given eigenvector
can be interpreted as the coupling strengths of the corresponding operators to
the state under consideration.

Our strategy differs from that of many previous studies that
utilized the variational method, in as far as our primary interest
lies in the couplings and not only in the resulting spectrum. We
first determine the eigenvalues of the three by three submatrices
within each of the Fock sectors, separately, in order to obtain an
``unperturbed'' spectrum. This is then used
to identify the affiliation of the eigenvectors to the eigenstates
on the jacknifes samples. Finally, we will compare spectrum and
eigenvector components
with the mixing elements switched on to this unmixed reference point.

\FIGURE{
\rotatebox{270}{\includegraphics[height=.9\textwidth,clip]{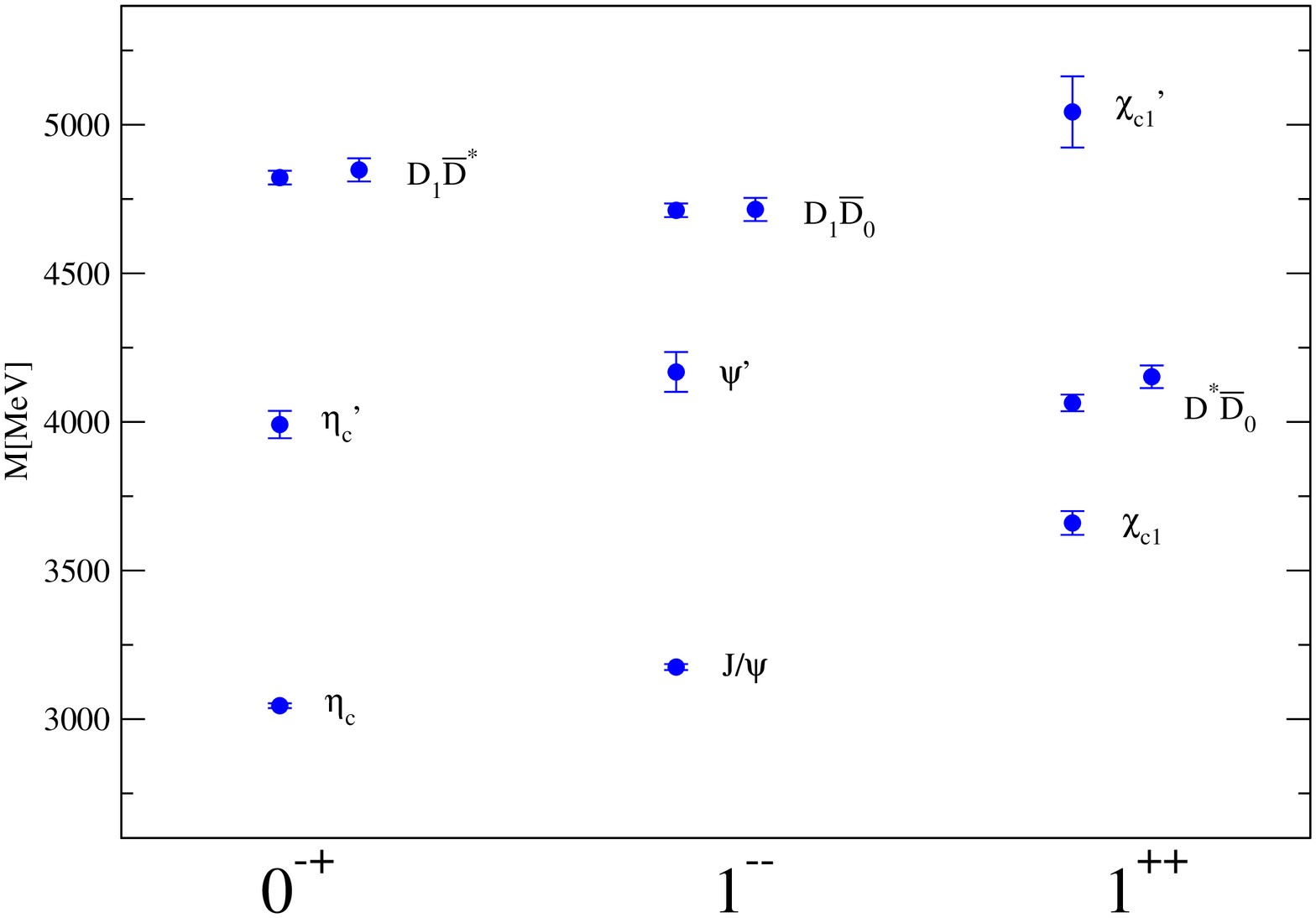}}
\caption{Mass spectra from the separate diagonalization of
the submatrices within each sector.
\label{spectrum_fig}}}
Our runs are performed on 100 effectively de-correlated $n_F=2$
$24^3\times 48$ configurations generated by the QCDSF
collaboration \cite{Gockeler:2008we}, with a lattice spacing
$a\approx 0.076\,$fm, obtained from the chirally extrapolated
nucleon mass. We employ the same non-perturbatively order $a$
improved clover action for the valence quarks. The charm quark
mass was set by tuning the spin-averaged charmonium mass
$\frac{1}{4}(m_{\eta_c}+3m_{J/\Psi})$
to its experimental value. The pion on these lattices is reasonably
light, $m_{\pi}\approx 280\,$MeV,
to embrace the above mentioned dependence of the mixing on the
light quark mass.

Computations were performed using the Chroma software
library \cite{Edwards:2004sx} partly on the local HPC cluster
and partly on the BlueGene/P of the J\"{u}lich Supercomputer Center. 

\section{Spectrum}
An extra benefit bucked off by our analysis
is the mass spectrum in the investigated channels.
The separate diagonalization of the three by three submatrices
provides us with at least four reliable eigenvalues, two for each
subsector. However, since the molecular channels typically are
rather noisy, we are only able to extract the ground states there,
within reasonable errors. So we are left with three states in each
channel, plotted in figure~\ref{spectrum_fig}.
For the molecular masses we give two data points:
the left ones are from the diagonalisation procedure,
the right ones represent the sums of the masses of the corresponding pairs of
non-interacting $D$ mesons. Note that in the $1^{++}$ channel,
the radially excited $\chi_{c1}$ is heavier than the molecular state,
in contrast to the other channels. If we consider the fact that our pion is
about 130-140 MeV too heavy, the mass of the molecular state is indeed
consistent with the $X(3872)$, which most likely has
$J^{PC}=1^{++}$ \cite{Abulencia:2006ma}.

\FIGURE
{
\rotatebox{270}{\includegraphics[height=.45\textwidth,clip]{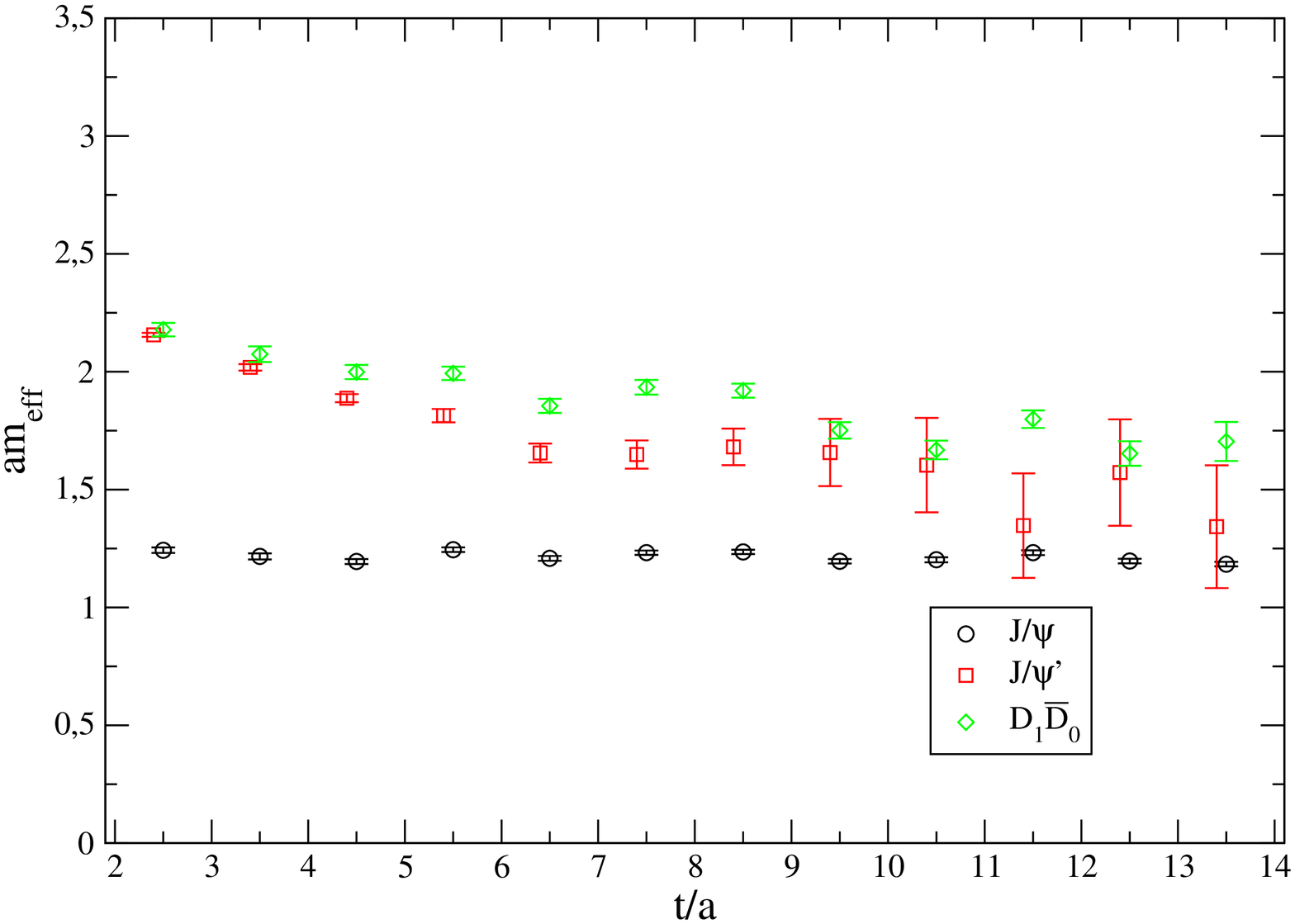}}\hspace*{.05\textwidth}
\rotatebox{270}{\includegraphics[height=.45\textwidth,clip]{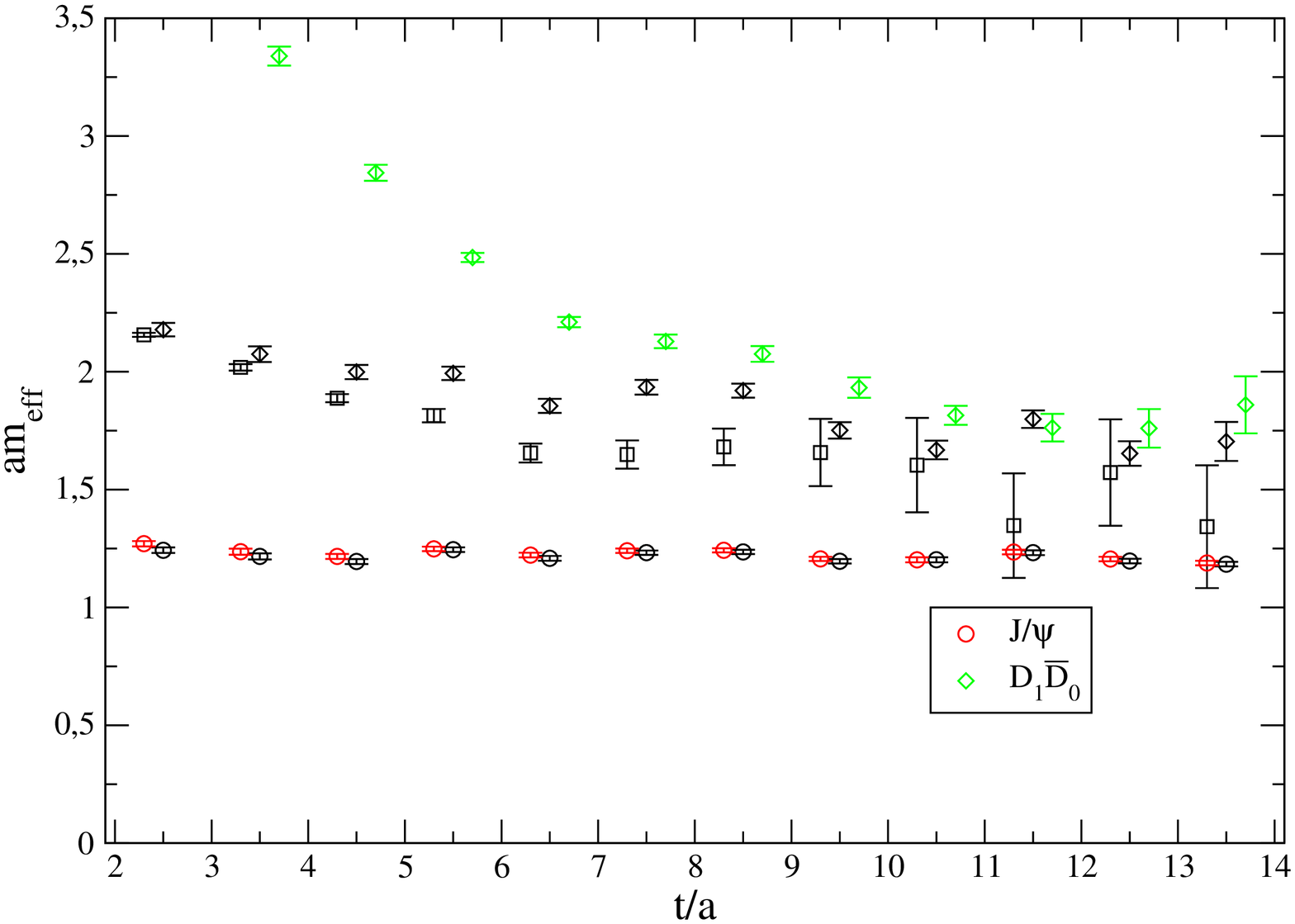}}
\caption{Effective masses of the eigenvalues of the submatrices (left hand side) and of the full matrix (right hand side) in the $1^{--}$ channel.
\label{eigenvalues_fig}}}

\section{Mixing}
Equipped with the reference eigenvalues from the submatrices, we go
for the diagonalization of the full six by six matrix. However, due to
limited statistics, we find
this to be numerically unstable
and restrict ourselves to the sub-basis
$M_{local},M_{narrow},Y_{local},Y_{narrow}$.
We discuss the vector state as one example.
In the left of figure~\ref{eigenvalues_fig} we display the effective
masses from the diagonalization of the two submatrices. The data
points for $J/\Psi$ and its radial excitation are from the $c\bar{c}$
submatrix, the ones for the $D_1\overline{D}_0$ from the molecular sector.

\FIGURE
{
\rotatebox{270}{\includegraphics[height=.45\textwidth,clip]{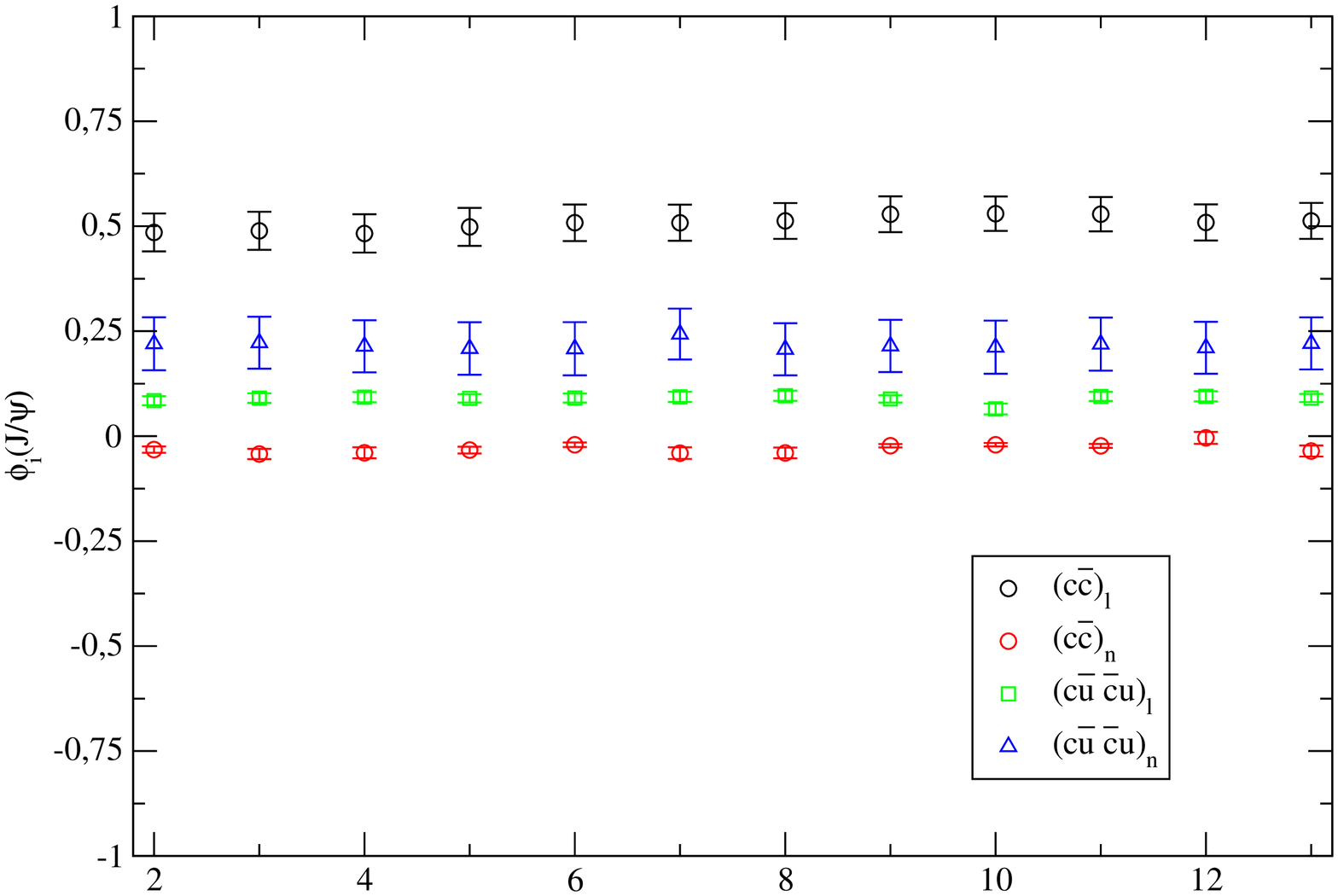}}\hspace*{.05\textwidth}
\rotatebox{270}{\includegraphics[height=.45\textwidth,clip]{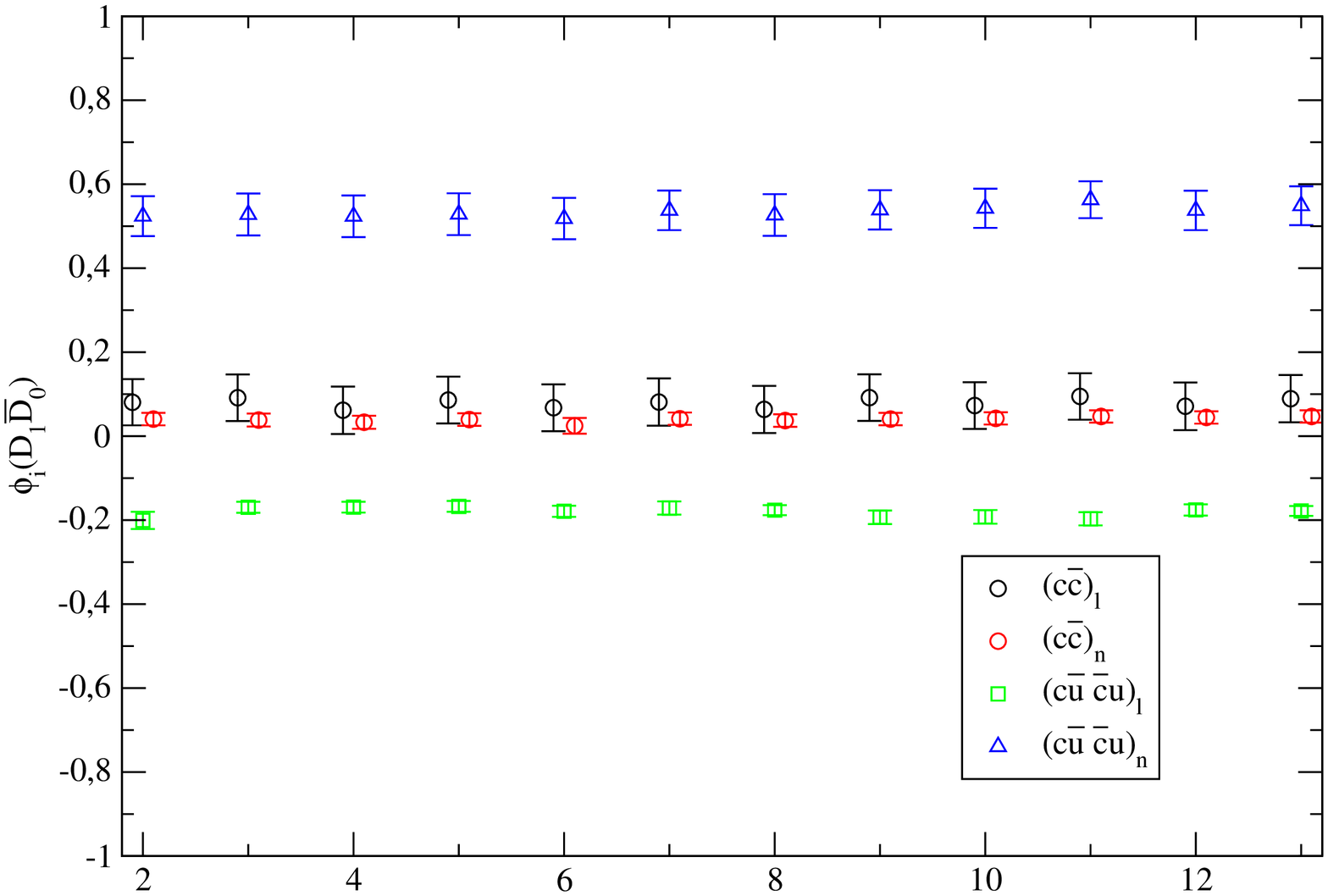}}
\caption{Eigenvector components from the diagonalization
of the full matrix in the $1^{--}$ channel.
\label{eigenvectors_fig}}}
\TABULAR{|c|c|c|c|c|}
{\hline
 state & $(c\bar{c})_l$ & $(c\bar{c})_n$ & $(c\bar{u}\bar{c}u)_l$ & $(c\bar{u}\bar{c}u)_n$ \\
 \hline
 $\eta_c$ & 0.54(3) & -0.02(1) & -0.1(1) & -0.31(5) \\
 $D_1\bar{D^*}$ & 0.07(1) & 0.01(1) & -0.46(8) & 0.14(2) \\
 \hline
 $J/\psi$ & 0.51(4) & -0.03(1) & 0.09(1) & 0.21(6) \\
  $D_1\bar{D}$ & 0.08(6) & 0.04(1) & -0.18(1)  & 0.53(4) \\
 \hline
 $\chi_{c1}$ & 0.39(5) & 0.69(3) & -0.22(3)  & -0.49(4) \\
  $D\bar{D^*}$ & 0.63(4) & -0.23(3) & -0.73(4) & 0.12(3) \\
 \hline}{Eigenvector components in the full basis.
\label{eigenveccomp_tab}}

The unmixed reference points can also be found in black color
in the right of
figure~\ref{eigenvalues_fig}. In addition,
the two lowest lying effective masses from the diagonalization
of the full matrix are shown there. We are able to identify these
two states with $J/\Psi$ and $D_1\bar{D_0}$, respectively.
Interestingly, the $J/\Psi'$ state is not found in the diagonalisation
of the four by four system.

The corresponding eigenvector components are plotted in
figure~\ref{eigenvectors_fig} for the $J/\psi$ (left) and
for the $D_1\overline{D}_0$ molecule.
The $J/\Psi$ receives the dominant contribution from the local
$c\bar{c}$ operator. However, the molecular configurations
seem to contribute significantly too. 
The $D_1\bar{D_0}$ state in contrast only contains
small (but non-vanishing) $c\bar{c}$ admixtures.

In table~\ref{eigenveccomp_tab} we summarize the results for
all channels that we investigated. In each of them we see significant
mixing effects. The large molecular contribution to the
$\chi_{c1}$ is particularly noteworthy to mention.

\section{Conclusion \& Outlook}
In all of the investigated channels ($0^{-+}$, $1^{--}$, and $1^{++}$)
we detect significant mixing effects between $c\bar{c}$ and four-quark
states. Although the precise values of the eigenvector components
should not be taken too seriously since the operator basis is rather small
and thus may miss non-negligible parts of the physical wavefunction,
our analysis clearly substantiates the assumption of charmonium states
having a rich Fock structure.

In the near future this study will be extended to other channels
including additional interpolating fields, e.g. a molecular operator
containing two mesons in a relative P-Wave.

\acknowledgments
This work was supported by the BMBF (contract 06RY257, GSI-Theory).
We thank the DFG Sonderforschungsbereich/Transregio 55 for their support.
We also thank the Forschungszentrum J\"{u}lich for providing
computer time on their Blue Gene/P system JUGENE and the QCDSF
collaboration for making their configurations available to us.

\end{document}